\documentclass[12pt,preprint]{aastex}

\begin{document}
\title{Hydrodynamic Collimation of Relativistic Outflows:  Semianalytic Solutions and Application to Gamma-Ray Bursts}
\author{Omer Bromberg and Amir Levinson\altaffilmark{1}}
\altaffiltext{1}{School of Physics and Astronomy, The Raymond and Beverly Sackler Faculty of Exact Sciences, Tel-Aviv University, Tel-Aviv 69978, Israel; Levinson@wise.tau.ac.il}
\begin{abstract}
A model is developed for the confinement and collimation of a baryon poor outflow by its surrounding medium.  Both, confinement by kinetic pressure of a static corona, and confinement by the ram pressure of a supersonic wind emanating from a disk surrounding the inner source are considered. Solutions are presented for the structure of the shocked layers of a deflected baryon poor jet (BPJ) and exterior wind. 
The dependence of the opening angle of the BPJ on the parameters of the confining medium are carefully examined.    
It is found that the BPJ shock may either converge to the symmetry axis or diverge away from it, depending on the opening angle of the BPJ injection cone.  In the latter case the inner flow exhibits a non-uniform structure, consisting of an ultra-relativistic core containing the unshocked BPJ enveloped by the slower, shocked BPJ layer. 
The implications of our results to the prompt GRB emission are briefly discussed.
\end{abstract}

\section {Introduction}


Collimation is a generic feature of outflows in astrophysics.  It appears in outflows from protostars, AGNs, microquasars and GRBs.
Despite many efforts to identify the mechanisms responsible for the acceleration and collimation of jets, there is as yet no universally 
accepted explanation for these processes.  One possibility is that jets are collimated by magnetic hoop stress that arises when the field is wound up by rotation.  This, however, requires ordered magnetic fields of sufficient strength to be present in the vicinity of the central engine.  
Whether such fields can be generated is yet an open issue.  Recent 3D simulations of accretion disks demonstrate that, although near equipartition magnetic fields can be generated by the MRI process, they are not expected to be ordered in general, except perhaps
within a flux tube that is connected to the horizon of a rapidly rotating black hole (McKinney \& Gammie 2004; McKinney, 2005; Hawley \& Krolik 2006; see also Krolik 2007, and references therein).  
Even if it can be produced, such a field configuration may be kink unstable (Lyubarsky 1992; Eichler, 1993; Begelman, 1998; but c.f., McKinney, 2006) and only slowly collimated (e.g., Begelman \& Li 1994, Lyutikov 2006; but c.f., Beskin \& Nokhrina 2006).

Another possibility is that the relativistic outflow is collimated by the pressure and inertia of a surrounding medium.  
A particular application of the latter mechanism is the prompt phase of GRBs, during which a multicomponent outflow 
is likely to be expelled by the central source:  The hyper-accretion rates anticipated following the collapse of a massive star or coalescence of compact objects render the disk surrounding the black 
hole extremely hot and dense.  Under such conditions the disk is optically thick to electromagnetic radiation 
and cools predominantly by emission of MeV neutrinos (Popham et al. 1999; Pruet et al. 2003) .
The prodigious neutrino luminosity drives a powerful, baryon rich wind from the inner disk radii that 
expands at sub or mildly relativistic speed.  The ultra-relativistic, baryon poor outflow is produced most likely in a polar region
which is protected from baryon contamination by either magnetic field lines threading the horizon or by an angular momentum barrier.
The energy that powers the inner, GRB-producing jet may be deposited by neutrino annihilation above the horizon (e.g., Levinson \& Eichler 1993; MacFadyen \& Woosley 1999; Aloy et al. 2000; Rosswog \& Ramirez-Ruiz 2003; Aloy et al. 2005).  Alternatively the energy may be extracted magnetically from the spin of the black hole (Levinson \& Eichler, 1993; Meszaros \& Rees, 1997; Van Putten, 2001; McKinney, 2005), or the inner disk regions (e.g., Vlahakis \& Konigl, 2003; Levinson, 2006a).  In this case the outflow is expected to be magnetically dominated, however, the magnetic energy should eventually be converted somehow to matter and radiation.
Collimation of the inner outflow by the surrounding wind is expected if the streamlines of the latter diverge slower than the streamlines of the former.

A related issue is the dissipation of the bulk energy of the fireball.
The standard view (see e.g., Piran 2005, and references therein), at least until recently, was that the prompt 
GRB emission is produced behind internal shocks that form in the coasting region of a baryon loaded fireball.
This scenario was originally motivated by the detection, in many sources, of a power law component
of the prompt emission spectrum extending up to energies well above the MeV peak,
in conflict with the thermal spectrum expected from an adiabatically expanding, baryon free 
fireball (Paczynski, 1986).  The post-SWIFT discoveries of a shallow afterglow phase at early times,
and fastly rising, large amplitude X-ray flares in the early afterglow phase
(Burrows et al. 2005) introduces new puzzles.  According to some interpretations (e.g., Fan and Wei 2005)
these observations imply prolonged activity of the central engine.  If true, then this may indicate that 
$\gamma$-rays are emitted during the prompt phase with very high efficiency, in the
sense that the remaining bulk energy is a small fraction of the total 
blast wave energy inferred from the afterglow emission.  Such episodes
can be quite naturally explained as resulting from dissipation in an electron-positron dominated plasma.
Alternative explanations involve variations of the micro-physical parameters and
do not require extremely high efficiency (Granot et al. 2006).

The difficulty with purely leptonic fireball models mentioned above may be alleviated if dissipation 
occurs during the outflow acceleration.  The claim that internal shocks must form in the coasting region,
since different shells cannot catch up at radii where the fireball
is accelerating, applies only to conical geometries.  As is shown below, collisions of the accelerating
fireball with the surrounding wind leads to formation of strong oblique shocks well below the coasting region.
Bulk Comptonization may then give rise to a nonthermal extension of the spectrum well above the thermal peak.
It is quite likely that these shocks pass through a photosphere
in which case the nonthermal photons may escape the system before being thermalized. 

A preliminary analysis of hydrodynamic collimation of a baryon poor outflow by a wind expelled from 
a thin torus has been presented in Levinson and Eichler (2000, hereafter LE00).  For this particular wind 
geometry LE00 have found that the opening angle equals the ratio of the power output of the inner fireball to that of the exterior baryonic wind, and suggested that the huge apparent isotropic luminosities implied for some GRB's (e.g., GRB 990123) can be understood in terms of collimation of the
fireball by an outer baryonic wind with more modest energetics, roughly
the geometric mean of the fireball energy and its isotropic equivalent. 
Detailed numerical simulations of acceleration and confinement of baryon free fireballs produced
by deposition of thermal energy above the poles of a non-rotating, stellar mass black hole that accretes matter
from a thick torus have been performed later by Aloy et al. (2005).   However, the confining medium in those simulations 
was restricted to some specific configurations obtained in NS+NS and NS+BH merger simulations, which might be relevant for
short GRBs.  Jet breakout in collapsars have been studied using axisymmetric hydrodynamic simulations (Aloy et al. 2000).  Magnetohydrodynamic simulations of outflow formation in collapsars have also been reported recently (Proga, et al. 2003).  

In this paper we generalize the model outlined in LE00 to incorporate additional important features that have 
been ignored there for the sake of simplicity.  In particular, we relax the assumption that the baryon poor jet (BPJ) is fully shocked and 
compute the structure of the shocked BPJ layer, explore other geometries of the confining medium, and consider 
also cases where the outer shock has a finite compression ratio.  Our approach is to construct semi-analytic models for the 
interaction of a baryon poor outflow with a surrounding medium, assuming steady state and axial symmetry.

\section {The model}\label{model}

We consider an accelerating, ultra-relativistic BPJ confined by the
pressure and inertia of a surrounding matter.  In case of confinement by a supersonic wind, we 
envision that the surrounding wind is expelled (e.g., from an extended disk or torus) over a
range of scales larger than the characteristic size of the central
engine ejecting the BPJ, so that not too far out its streamlines may
diverge more slowly than the BPJ streamlines, thereby giving rise to a
collision of the two outflows.  In general, such a collision will lead to
the formation of a contact discontinuity across which the total pressure
is continuous, and two oblique shocks, one in each fluid, across which
the streamlines of the colliding (unshocked) fluids are deflected, as shown schematically in fig 1.
The details of this structure will depend, quite generally, on the parameters of
the two outflows and on the boundary conditions.

Since we are merely interested in the ejection phase following the outflow formation in the case of coalescence of a compact binary,
or the breakout episode in the case of explosion inside a star, as in the collapsar model, it is reasonable to assume 
that the system under consideration is time independent and axially symmetric, which greatly simplifies 
the analysis.  This of course excludes any temporal effects 
associated with Kelvin-Helmholtz instabilities, mixing of the two fluids, etc., at the interface separating the two outflows  which 
can only be studied using 3D numerical simulations.  On the other hand, it provides a simple and convenient way to 
study the gross structure of such a system and to gain some insight into the physics involved.  It may also provide a test
case for more sophisticated simulations.  To this end, we construct in what follows semi-analytic models 
of colliding outflows.

\subsection {Basic equations}\label{equations}

We seek to determine the structure of the shocked layers assuming that
the parameters of the unshocked fluids are given.  
The problem is then characterized by the following independent variables:
the rest mass density $\rho_{js}$, pressure $p_{js}$, specific enthalpy $w_{js}$,
and 4-velocity $u^\nu_{js}$ of the shocked BPJ fluid;
the rest mass density $\rho_{ws}$, pressure $p_{ws}$, specific enthalpy $w_{ws}$,
and 4-velocity $u^\nu_{ws}$ of the shocked wind fluid, and the cross-sectional radii of the
inner shock surface $r_j(z)$, outer shock surface $r_w(z)$ and the contact discontinuity surface $r_c(z)$.
(Henceforth, subscripts $j$, $js$, $w$ and $ws$ refer to quantities in the unshocked inner jet, in the shocked jet layer,
in the surrounding wind, and in the shocked wind layer, respectively.) The various zones are indicated in fig 1.
The stress-energy tensor associated with the inner (outer) shocked layers can be expressed as
\begin{equation}
T^{\mu\nu}_{js(ws)} = w_{js(ws)}u_{js(ws)}^{\mu}u_{js(ws)}^{\nu} + p_{js(ws)}g^{\mu\nu},
\label{T_M}
\end{equation}
where $g^{\mu\nu}$ is the metric tensor.  
In the what follows we use cylindrical coordinates ($r,\phi,z$), and assume that the system is time independent and axially symmetric.   The 
dynamics of the system is governed by mass conservation in each shocked layer: 
\begin {eqnarray}
\partial_k (\rho_{js} u^k_{js}) &=& 0,\label{cont-j}\\
\partial_k (\rho_{ws} u^k_{ws}) &=& 0,\label{cont-w}
\end{eqnarray}
where the index $k$ in the above equations runs through $r$ and $z$,
energy-momentum conservation:
\begin {eqnarray}
\partial_k  T_{js}^{\mu k} &=& 0,\label{T-j}\\
\partial_k  T_{ws}^{\mu k} &=& 0,\label{T-w}
\end{eqnarray}
and an equation of state for the fluids in each shocked layer.
The above system of equations is subject to boundary conditions at the shock and contact discontinuity 
surfaces, obtained from the jump conditions across each surface. Continuity of energy, momentum and mass across 
the inner shock surface implies
\begin {eqnarray}
(T_{js}^{\mu\nu}n_{j\nu})_{r=r_j} &=& (T_{j}^{\mu\nu}n_{j\nu})_{r=r_j},\label{jump-js1}\\
(\rho_{js} u^\nu_{js}n_{j\nu})_{r=r_j} &=& (\rho_{j} u^\nu_{j}n_{j\nu})_{r=r_j}\label{jump-js2},
\end{eqnarray}
where $n_{j\nu}=(0,\hat{n}_j)$ is a space-like unit vector normal to the shock front.
Likewise, for the outer shock surface we have
\begin {eqnarray}
(T_{ws}^{\mu\nu}n_{w\nu})_{r=r_w} &=& (T_{w}^{\mu\nu}n_{w\nu})_{r=r_w},\label{jump-ws1}\\
(\rho_{ws} u^\nu_{ws}n_{w\nu})_{r=r_w} &=& (\rho_{w} u^\nu_{w}n_{w\nu})_{r=r_w}.\label{jump-ws2}
\end{eqnarray}
Finally, at the contact discontinuity the streamlines of the shocked fluids must be tangent to the surface.
The jump conditions there then reduce to
\begin{eqnarray}
(u_{js}^\mu n_{c\mu})_{r=r_c}=(u_{ws}^\mu n_{c\mu})_{r=r_c}=0,\label{jump-c1}\\
(p_{js}-p_{ws})_{r=r_c}=0,\label{jump-c2}
\end{eqnarray}
where $n_{c\mu}=(0,\hat{n}_c)$ is a unit vector normal to the contact discontinuity surface.

It is convenient to express the terms on the right hand side of eqs. [\ref{jump-js1}]- [\ref{jump-ws2}] 
explicitly in terms of the parameters of the unshocked BPJ and wind fluids, which are given as input, and the 
impact angles $\delta_{j}$ and $\delta_w$ (that is, the angle between the velocity of the unshocked jet/wind fluid 
and the corresponding shock surface), which are unknown a priori.  Denoting by $u_{j(w)}=(u_{j(w)}^zu_{j(w)z}+u_{j(w)}^ru_{j(w)r})^{1/2}$ the poloidal velocity of the unshocked jet (wind) and by $\gamma_{j(w)}=(1+u_{j(w)}^2)^{1/2}$
the corresponding Lorentz factor, the mass and energy fluxes that pass through the inner (outer) shock can be written as
\begin {eqnarray}
(\rho_{j(w)} u^\nu_{j(w)}n_{j(w)\nu})_{r=r_{j(w)}} &=& \rho_{j(w)} u_{j(w)}\sin\delta_{j(w)},\label{def-delta1}\\
(T_{j(w)}^{0\nu}n_{j(w)\nu})_{r=r_{j(w)}} &=& w_{j(w)}\gamma_{j(w)}u_{j(w)}\sin\delta_{j(w)}.\label{def-delta2}
\end{eqnarray}
The flux of transverse momentum incident into the inner (outer) shock can be obtained by projecting the 
RHS of eq. (\ref{jump-js1}) (eq. [\ref{jump-ws1}]) on the direction normal to 
the shock front:
\begin {equation}
(T_{j(w)}^{\mu\nu}n_{j(w)\mu}n_{j(w)\nu})_{r=r_{j(w)}} = w_{j(w)}u^2_{j(w)}\sin^2\delta_{j(w)}+p_{j(w)}.
\label{p-flux-jn}
\end{equation}
Finally, projecting eqs. (\ref{jump-js1}) and (\ref{jump-ws1}) on the axial direction we obtain 
\begin{equation}
(T_{j(w)}^{z\nu}n_{j(w)\nu})_{r=r_{j(w)}} = w_{j(w)}u^2_{j(w)}\cos\theta_{j(w)}\sin\delta_{j(w)},
\end{equation}
where $\theta_{j(w)}$ denotes the angle between the velocity of the unshocked BPJ (wind) and the symmetry 
axis at the shock front.  The angles $\theta_j$ and $\theta_w$ depend in general on the geometry of streamlines.  For the 
BPJ we assume conical streamlines, so that to a good approximation $\tan\theta_j=r_j/z$.  The impact
angle is related to $r_j(z)$ through
\begin{equation}
\sin\delta_j=\frac{(\sin\theta_j-\cos\theta_j r_j^\prime)}{(1+r_j^{\prime 2})^{1/2}}
={(r_{j}-r_{j}^\prime z)\over [(r_{j}^2+z^2)(1+r_{j}^{\prime 2})]^{1/2}},
\label{delta_j}
\end{equation}
where $r_j^\prime\equiv dr_j/dz$. Likewise, for the outer wind the relation between $\delta_w$, $\theta_w$ and $r_w$ 
is obtained upon substituting $r_j\rightarrow r_w$, $\theta_j\rightarrow \theta_w$ and $\delta_j\rightarrow -\delta_w$ in eq. (\ref{delta_j}).  The angle $\theta_w(z)$ is computed numerically for each of the flow geometries considered below.

Under the conditions anticipated the shocked BPJ plasma is radiation dominated.  The equation of state of the shocked BPJ fluid 
can be approximated as,
\begin{equation}
w_{js}=4p_{js}+\rho_{js}c^2.\label{state_eq_jet}
\end{equation}
In the shocked wind layer the pressure contributed by baryons may be important.
The total pressure there is then the sum of the radiation pressure 
and the pressure contributed by protons and electrons:
\begin{equation}
p_{ws}=a T_{ws}^4+2\frac{\rho_{ws}}{m_p}kT_{ws},\label{pressure_wind}
\end{equation}
where $T_{ws}$ is the temperature of the shocked gas, and it is assumed that $kT_{ws}/m_ec^2$ is typically 
less than unity, so that the electrons can be considered non-relativistic.  The corresponding specific enthalpy is given by, 
\begin{equation}
w_{ws}=\rho_{js}c^2+4aT_{ws}^4+5\frac{\rho_{ws}}{m_p}kT_{ws}.\label{state_eq_wind}
\end{equation}

To simplify the analysis further, we suppose that each shocked layer can be approximated as a
one dimensional, cylindrically symmetric flow along the channel.  Specifically, it is assumed that 
inside the shocked layers the flow parameters depend solely on $z$.  Equations (\ref{cont-j})-(\ref{T-w}) can then be integrated 
over a section of the flow located between $z$ and $z+dz$.    Moreover, as shown in the appendix in 
the small angle approximation, more precisely to order $O(r^{\prime2})$, the poloidal 4-velocity equals the $z$ component of the 4-velocity, viz., $u_{js}=u^z_{js}$ and likewise  for $u_{ws}$. By employing the latter result we can eliminate two variables thereby obtaining a closed set.  The details are given in the appendix.  Using this procedure, we obtain from eqs. (\ref{cont-j}) and (\ref{cont-w}) mass conservation laws for the deflected BPJ and wind fluids:
\begin{eqnarray}
&&\frac{d}{dz}\left[\rho_{js}u_{js} (r_c^2-r_j^2)\right] =
2\rho_ju_jr_j\frac{\sin\delta_j}{\cos\alpha_j},\label{mass_eq_j}\\
&&\frac{d}{dz}\left[\rho_{ws}u_{ws} (r_w^2-r_c^2)\right] =
2\rho_wu_wr_w\frac{\sin\delta_w}{\cos\alpha_w}.\label{mass_eq_w}
\end{eqnarray}
Likewise, energy conservation is obtained from the zeroth component of eqs. (\ref{T-j}), (\ref{T-w}):
\begin{eqnarray}
&&\frac{d}{dz}\left[w_{js}\gamma_{js}u_{js}(r_c^2-r_j^2)\right] =
2w_j\gamma_ju_jr_j\frac{\sin\delta_j}{\cos\alpha_j},\label{energy_eq_j}\\
&&\frac{d}{dz}\left[w_{ws}\gamma_{ws}u_{ws}(r_w^2-r_c^2)\right] =
2w_w\gamma_wu_wr_w\frac{\sin\delta_w}{\cos\alpha_w},\label{energy_eq_w}
\end{eqnarray}
and momentum conservation from the $z$ component of eqs. (\ref{T-j}) and (\ref{T-w}):
\begin{eqnarray}
&&\frac{d}{dz}\left[w_{js}u_{js}^2(r_c^2-r_j^2)-r_j^2p_{js}\right]+r_c^2{dp_{js}\over dz}+p_j{dr^2_j\over dz} = 
2w_ju_j^2r_j\frac{\cos\theta_j\sin\delta_j}{\cos\alpha_j},\label{momZ_eq_j}\\
&&\frac{d}{dz}\left[w_{ws}u_{ws}^2(r_w^2-r_c^2)+r_w^2p_{ws}\right]-r_c^2{dp_{ws}\over dz}- p_w{dr^2_w\over dz} = 
2w_wu_w^2r_w\frac{\cos\theta_w\sin\delta_w}{\cos\alpha_w}.\label{momZ_eq_w}
\end{eqnarray}
Projecting eqs. (\ref{T-j}) and (\ref{T-w}) on the direction normal to the inner and outer shock fronts, respectively, we obtain
(to order O$(r^{\prime2})$, see appendix) the conditions
\begin{eqnarray}
&&p_{js}=w_ju_j^2\sin^2\delta_j+p_j,\label{momN3_eq_j}\\
&&p_{ws}=w_wu_w^2\sin^2\delta_w+p_w.\label{momN3t_eq_w}
\end{eqnarray}
Finally, eq. (\ref{jump-c2}) implies equal pressures in the two shocked layers, viz.,
\begin{equation}
p_{js}=p_{ws}\label{p_eq_c}.
\end{equation}
The system of equations (\ref{state_eq_jet})-(\ref{p_eq_c}) forms a closed set for the 11 independent variables that characterize the shocked BPJ and wind fluids.

\subsection{Input parameters and boundary conditions}
The BPJ is assumed to be injected from a point source into a cone of opening angle $\theta_{j,max}$,
with a uniform energy distribution inside the cone.  The exterior wind is envisioned to be ejected from a thin disk located 
at the equatorial plane of the system.  We consider separately subsonic winds (or coronae), for which confinement is dominated 
by kinetic pressure, and supersonic winds for which kinetic pressure can be ignored.  The situation of a static gas column
with a given pressure profile (e.g., the envelope of a star) is treated as a special case of such a subsonic wind. 
As our canonical model we adopt a split monopole geometry for the streamlines of the confining wind (see fig 2), which captures most of 
the basic features in a convenient way, but examine also other wind geometries for comparison.
The wind is assumed to be uniformly ejected over the range of angles $\theta_{w,min}\le\theta_w\le \pi/2$ (in the upper half plan),
and has a total luminosity $L_w$.
The integration starts at the position where collision of streamlines first occurs (see fig 2).  We find it most 
convenient to obtain the initial values of the flow quantities in the shocked layers using the jump conditions of
an infinite planar, oblique shock.  The direction of the shock surface at the origin is obtained as an eigenvalue of the corresponding 
system of algebraic equations.   To check the sensitivity of our solutions to the initial values, we made some runs with different 
initial conditions.  In most cases it has been found that the solution is highly insensitive to boundary conditions, and that after some 
initial phase the integration quickly converges to the same solution.

\section{Results}\label{test&results}

\subsection{Pressure confinement}

We consider first the possibility that the BPJ is collimated by the kinetic 
pressure of the surrounding medium.   
As mentioned above, this is naturally expected when the surrounding medium
is static or highly subsonic, as in the case where the confining medium
is the stellar envelope into which a fireball is ejected in the collapsar model for GRBs.
This is also expected when the confining medium is mildly supersonic if
$\delta_w<{\cal M}_w^{-1}$, where ${\cal M}_w$ is the Mach number of the enveloping wind.
We envisage that the confining medium has a geometry of a torus.  To be more concrete,
we suppose that within some cylindrical radius $R$ (of the order of a few Schwarzchild radii) the region near the injection point of the BPJ is devoid of matter, so that collision of the BPJ with the enveloping corona first occurs at a height $z_0=R/\tan\theta_{j,max}$ above the equatorial plan.  This choice of boundary conditions merely reflects the expected configuration after the initial transient phase (e.g., the corking of the stellar envelope in the collapsar model).
Formally, the limit of kinetic pressure domination is obtained in our model upon substituting $u_w=0$, $r_w=r_c$ in 
eqs. (\ref{mass_eq_j})-(\ref{momN3t_eq_w}).  Equations (\ref{momN3t_eq_w}) and (\ref{p_eq_c}) then imply 
$p_{js}=p_w$.  For convenience, we adopt a power law profile for the external pressure, 
$p_{w}\propto z^{-\eta}$.  When $\eta<4$ the oblique shock that forms as a result of collision of the 
BPJ with the confining walls should diverge slower than conical or even 
converge to the axis.  This can be readily understood by noting that if the shock surface is a cone (i.e., $r_{j}=R+\tan\alpha_j z$
with $\alpha_j=$ const), 
then the impact angle (eq. [\ref{delta_j}]) decreases as
\begin{equation}
\sin\delta_j={(r_{j}-r_{j}^\prime z)\over [(r_{j}^2+z^2)(1+r_{j}^{\prime 2})]^{1/2}}\simeq {R\over z}
\end{equation}
at $z>>R$, and the pressure inside the shocked layer scales as (see eq. [\ref{momN3_eq_j}])
\begin{equation}
p_{js}\simeq w_ju_j^2 \sin^2\delta_{j}= \frac{L_{j}\sin^2\delta_{j}}{2\pi(1-\cos\theta_{j,max})(r_{j}^2+z^2)}\propto z^{-4},
\end{equation}
where $L_j$ is the total power and $\theta_{j,max}$ is the opening angle of the unshocked jet at the injection point.
The requirement that the pressure is continuous across the contact discontinuity surface, $p_w=p_{js}$, implies
in turn $\eta=4$.  Evidently, when $\eta<4$ the shock surface cannot be conical\footnote{Using similar arguments Eichler (1982) has shown that if the shocked BPJ fluid cools rapidly, then for $\eta=4$ the cross-sectional radius of the BPJ is independent of $z$, that is, the BPJ is confined in a cylinder.}. 
An example is shown in fig 3, where results of numerical integration are exhibited for $p_w=p_{w0}(z/z_0)^{-3}$ and two different values of $\theta_{j,max}$.  In both cases shown the contact discontinuity surface 
has the same shape.  The evolution of the inner shock surface, however, depends on $\theta_{j,max}$, as can be seen.
Whether the shock front converges to the axis or diverges away from it depends on the value of the impact angle required 
to sustain momentum balance (see also Komissarov \& Falle, 1997).   We find that when the condition
$p_{w0}>L_j\sin^2 2\theta_{j,max}/[16\pi c R^2(1-\cos\theta_{j,max})]$ is satisfied, the shock front is inclined towards the axis and the BPJ quickly becomes fully shocked.  The reflection of the oblique shocks at the symmetry axis should lead to formation of internal shocks, which may affect somewhat the resultant structure (e.g., Aloy et al. 2000).   In particular it is expected to lead to inhomogeneities in the parameters of the shocked BPJ.  
However, we believe that this should not affect the average structure dramatically.

In the region where the BPJ can be considered fully shocked we have $r_j=0$.  Equations (\ref{energy_eq_j}) and  (\ref{momZ_eq_j}) 
then reduce to
\begin{equation}
4p_{js}\gamma_{js}u_{js}\pi r_c^2=L_j,\label{energy-kp}
\end{equation}
and
\begin{equation}
\frac{d}{dz} \ln\gamma+\frac{1}{4}\frac{d}{dz} \ln p_{js}=0,\label{mom-kp}
\end{equation}
where $w_{js}=4p_{js}$ has been adopted.  Solving eqs (\ref{energy-kp}) and (\ref{mom-kp}) 
and using the condition $p_{js}=p_w(z)$ we recover the result obtained by LE00,
\begin{equation}\label{r_c-Pressure}
r_c(z)=r_{co}[p_w(z)/p_{wo}]^{-1/4}\propto z^{\eta/4},
\end{equation}
which holds in the BPJ acceleration zone.  Evidently, collimation requires $\eta<4$.  As pointed out in LE00, in
the case where the pressure is contributed by a transonic wind, such
collimation will occur naturally in the region located within the
acceleration zone of the BPJ and above the acceleration zone of the
surrounding wind (roughly above its critical point), even if the two
outflows have the same equation of state.  The reason is that in this
region the density profile of the BPJ declines with radius as $r^{-3}$
(for a conical geometry) while that of the slow wind declines as
$r^{-2}$, since the wind has reached its terminal velocity.  Adopting
an equation of state of the form $p_w(n_w)\propto n_w^{\Gamma}$ for
the baryonic fluid yields $\eta=2\Gamma$ for a conical (or spherical)
wind, for which $n_w(z)\propto z^{-2}$.  Convergence occurs for
$\Gamma<2$, which is the case for both relativistic gas ($\Gamma=4/3$)
and non-relativistic gas ($\Gamma=5/3$).
The BPJ breakout angle from a collapsar surface can be estimated from equation (\ref{r_c-Pressure}) to be,
\begin{equation}\label{alpha*}
\alpha_{c*}\simeq r'_c(z_*)=\kappa(r_{co}/z_o)(z_{*}/z_o)^{\kappa-1},
\end{equation}
where $\kappa=\eta/4$ if the coasting radius lies outside the stellar envelope and $\kappa=\eta/2$
if it lies inside, $z_*\sim 10^{10}$ cm is the location of the stellar envelope (e.g MacFadyen \& Woosley, 1999).
For example in the configuration shown in fig 3 we find that
$\alpha_*=0.04~{\rm rad}$. 

It is worth noting that collimation shocks should continue to form near the interface separating the two outflows whenever deflection
of streamlines occurs.  Such effects cannot be accounted for by
the 1D model presented here.

\subsection{Confinement by a supersonic wind}\label{res_ram_torus}

We consider next the situation wherein the BPJ collides with a supersonic wind.  The analysis can be simplified
by assuming that the exterior wind is deflected into a very thin shock compared with the cross-sectional radius of the contact
discontinuity, and that on all scales the momentum transfer is dominated by the ram pressure of the wind.  
In this limit one can take $r_w=r_c$ in the above equations.  This effectively assumes an infinite compression ratio for the 
outer shock and may be a reasonable
approximation when the cooling time of the shocked fluid is much shorter than the expansion time. However, 
under conditions anticipated in GRBs the optical depth of the shocked wind layer is expected to be very large 
on scales of interest and radiative cooling can be ignored.  As is shown below, the resultant pressure of the shocked wind
layer is larger in this case, owing to the larger inclination of the shock front, thereby giving rise to a better collimation.
Nonetheless, for illustrative purposes we analyze first the case where the deflected wind layer is negligibly thin.  With $r_c=r_w$ equations
(\ref{state_eq_jet}), (\ref{mass_eq_j}), (\ref{energy_eq_j}), and (\ref{momZ_eq_j}), and the condition
\begin{equation}
w_ju_j^2\sin^2\delta_j+p_j=w_wu_w^2\sin^2\delta_w+p_w, \label{jump-rad-c}
\end{equation}
which follows from eqs. (\ref{momN3_eq_j})-(\ref{p_eq_c}), 
provide a complete description of the BPJ structure.

Further simplification is possible when the inner shock angle is sufficiently small, viz., $\alpha_j<<\alpha_c$, such that 
the energy and momentum fluxes incident through the inner
shock can be neglected.  Upon taking $\delta_j=0$, the system of 
equations (\ref{state_eq_jet}), (\ref{mass_eq_j}), (\ref{energy_eq_j}), (\ref{momZ_eq_j}), and (\ref{jump-rad-c})
reduce to a single, nonlinear differential equation for the cross-sectional radius $r_c(z)$ (LE00):
\begin{equation}
(r_c/r_{c0})^{-4}=\frac{w_w u_w^2}{p_{js,0}}\frac{(\sin\theta_w-\cos\theta_w r_c^\prime)^2}{(1+r_c^{\prime 2})}.
\label{cross-sec-rad}
\end{equation}
Here $r_{c0}=r_c(z=z_0)$ and $p_{js,0}=p_{js}(z=z_0)$ are the cross-sectional radius and pressure at some 
fiducial point $z=z_0$.  Solutions of the latter equation can be obtained once the exterior wind configuration 
is specified.  LE00 solved eq. (\ref{cross-sec-rad}) in the case of a wind emanating from a thin 
torus of radius $R_T$.  They have found that at $z>R_T$ the shocked jet is confined in a cone of 
opening angle $\alpha_c\simeq \chi^{-1}$,
where $\chi=L_w/L_j$ is the ratio of the wind and jet luminosities.  For the split monopole wind depicted in fig 2 we have $\tan\theta_w=r_c/(z+d)$ and 
$w_w u_w^24\pi c\cos\theta_{w,min}[r_c^2+(z+d)^2]=L_w$.  Equation (\ref{cross-sec-rad}) then yields $r_c=r_{c0}+\tan\alpha_{c0}(z-z_0)$ at 
$z\ge z_0>>d$, where $\alpha_{c0}$ denotes the value of $\alpha_c(z)$ at $z=z_0$, with some nontrivial dependence 
of $\alpha_{c0}$ on $\chi$.  Of course $\alpha_{c0}$ is limited by geometry, that is,
$\alpha_{c0}$ cannot be smaller than $\theta_{w,min}$.  It can be readily shown that this behaviour is not limited to the two examples discussed above; 
for any wind configuration that 
satisfies $\tan\theta_w\simeq r_c/z$ and $w_w u_w^2 \propto z^{-2}$ at large $z$, the shocked jet becomes conical, with 
an opening angle that depends on the geometry and power of the unshocked wind.  

The above trend have been confirmed by numerical integration of the full set of 
equations (\ref{state_eq_jet}), (\ref{mass_eq_j}), (\ref{energy_eq_j}), (\ref{momZ_eq_j}), and (\ref{jump-rad-c}).
An example is shown in fig 4 for the split monopole wind exhibited in fig 2 with $\theta_{w,min}=8.5^\circ$.  
The inner jet in this example has an initial opening angle of $\theta_{j,max}=45^\circ$, and its power is normalized such that $L_j/4\pi c R^2=2\times10^{26}$ ergs cm$^{-3}$, where $R=d\tan\theta_{w,min}$ is the radius at which the innermost streamline of the split monopole wind intersects the equatorial plane (see fig 2).  Collimation occurs typically when $\chi$ exceeds unity.
We find that for values of $\chi$ not too large ($\chi<200$ for the choice of parameters in fig 4) the inner shock diverges away from the axis.  The coaxial structure that forms then consists of an inner, fast core (the unshocked jet), and a slower sheath surrounding it (the shocked layer), as seen in the left panels in fig 4.
For very large values of $\chi$ the shock surface converges towards the axis and the inner jet becomes fully 
shocked, as seen in the right panels in fig 4.  At sufficiently large radii the contact discontinuity surface
indeed becomes conical as expected, with an opening angle that scales roughly as $\alpha_{c0}\propto \chi^{1/3}$ at $\alpha_{c0}>\theta_{w,min}$.  
The dependence of $\alpha_{c0}$ on the power ratio $\chi$ is presented in fig 5.  The above results depend only weakly on the 
opening angle of the BPJ injection cone $\theta_{j,max}$.  Changing the latter from $45^\circ$ to $60^\circ$ altered the above results by no more than
a few percent.

A stronger dependence of $\alpha_{c0}$ on $\chi$ has been found for exterior wind configurations in which some streamlines are
inclined towards the axis, as in the case of a wind from a thin torus invoked in LE00, where $\alpha_{c0}\propto\chi^{-1}$.  This stronger dependence is a consequence of the large impact angles $\delta_w$ that give rise to a much stronger compression near the injection point of the BPJ.

Better collimation can be achieved when the streamlines of the exterior wind diverge slower than conical.
An example is a parabolic wind, for which the family of streamlines is defined by the equation
\begin{eqnarray}\label{parabolic_coor}
\Psi(r,z)=\sqrt{z^2+r^2}-z=r_0,
\end{eqnarray}
with $r_0$ being the family parameter.  A particular choice of $r_0$ corresponds to the radius at which a given 
streamline meets the disk.  The impact angle is given by 
\begin{equation}\label{tanT}
\tan\theta_w=\frac{r_0}{r}=\frac{\sqrt{z^2+r^2}-z}{r},
\end{equation}
which at $z>>r_0$ satisfies, $\tan\theta_w\rightarrow r/(2 z)$.  
The power enclosed between two neighboring surfaces $\Psi=r_0$ and $\Psi=r_0+dr_0$:
\begin{equation}\label{dL}
dL=w_wu_w^22\pi r dr\cos\theta_w=w_wu_w^2 2\pi r^2\frac{\sqrt{dr_0^2+2dr_0 z}}{\sqrt{r^2+r_0^2}}
\end{equation}
is conserved.  At $z>>r_0$ the last equation yields $dL\simeq w_wu_w^24\pi z\sqrt{r_0 dr_0}$, implying $w_wu^2_w\propto z^{-1}$.  Assuming again that the BPJ is fully shocked, substituting the above results into eq. (\ref{cross-sec-rad}) and solving for $r_c(z)$, one readily finds $r_c\propto \gamma\propto z^{1/2}$, and $p_{js}\propto z^{-2}$.
Integration of the full set (\ref{state_eq_jet}), (\ref{mass_eq_j}), (\ref{energy_eq_j}), (\ref{momZ_eq_j}) and (\ref{jump-rad-c}) confirms
again this behaviour.  Because of the larger flux of transverse momentum through the outer shock, convergence of the inner shock towards the axis occurs at much lower values of $\chi$ than in the split monopole case.

We now consider the limit where energy losses by the deflected wind layer are negligible.   In that case the deflected wind layer cannot be considered thin and the constraint $r_w=r_c$ must be relaxed.  We must then numerically integrate the full set of equations (\ref{state_eq_jet})-(\ref{p_eq_c}). 
As explained above, eqs. (\ref{state_eq_jet})-(\ref{p_eq_c}) have been derived using the small angle approximation.  However, for the cases examined below we find that the angle of the outer shock surface, $\alpha_w$, is typically not small.  This leads to an overestimate of the ratio of 
energy and momentum fluxes incident through the shock front, and tends to produce some artificial dissipation inside the shocked wind layer.  To correct for this effect, we computed the energy and momentum fluxes that pass through the shock locally by solving the jump conditions of an infinite planar oblique shock.  Results of such calculations, using for a comparison a split monopole wind with the same parameters used
in fig 4, are exhibited in fig 6.  Comparison of figs 4 and 6 reveals that better collimation is achieved in the absence of energy losses behind the shock.  The BPJ in this case does not become conical asymptotically, as in the thin shock case exhibited in fig 6, 
but rather the profile of the contact discontinuity surface approaches $r_c\propto z^{2/3}$ roughly.
The reason for this behaviour is that the smaller compression ratio of the shock gives rise to a larger inclination of the shock front and, hence, a larger impact angle $\delta_w$, as seen in fig 6.  This in turn leads to a larger pressure behind the shock, as implied by eq. (\ref{momN3t_eq_w}).  
  
Our analysis does not take into account effects associated with the formation of a rarefaction wave near the interface separating the shocked BPJ and wind fluids in the region where the deflected wind layer is highly supersonic.  This should
give rise to a somewhat steeper decline of the pressure supporting the BPJ, and the consequent alteration of the resultant structure 
in this region.  Under certain conditions the fluid that resides between the contact discontinuity and the rarefaction wave can quickly accelerate to high Lorentz factors (Rezzolla et al. 2003; Aloy \& Rezzolla 2006).
We find that the contact discontinuity surface becomes causally disconnected from the outer shock front (
that is, the shocked wind layer expands sideways at a speed that exceeds the local sound speed) at $z/R>$ a few.  At this height the BPJ has already been substantially collimated.  For $\chi=18$ and $\chi=576$ we find $\alpha_c\simeq12^\circ$ and $6^\circ$, respectively at the decoupling point.

\section{Limitations of the Model}\label{limit}
The main limitation of our model is that it cannot account for any
temporal effects, particularly those associated with Kelvin-Helmholtz
instabilities at the interface separating the two outflows.  Linear
analysis by Hardee \& Hughes (2003) for a cylindrical, non accelerating
flow embedded in an external medium indicates that non-axisymmetric
modes, particularly low order helical and elliptical modes,
are likely do grow on time scales of the order of the jet crossing time. The
presence of a viscous boundary layer (that they modeled as a wind) leads to 
suppression of the non-axisymmetric modes and an increase of the
growth rate of the zeroth order axisymmetric pinch mode. If the
instability grows to a nonlinear state over the expansion time then
pinching of the fast jet and mixing of wind material with the light
jet fluid near the interface is anticipated.  This can modify the
transverse structure of the jet, and produce internal shocks that can
lead to further dissipation of the bulk energy of the jet (see next
section for further discussion).  Numerical simulations exhibit some
evidence for such instabilities (e.g., Aloy, 2005; Hardee \& Hughes
2003), although the details may depend on the configuration of the
confining medium, and on the structure of the viscous boundary layer.
In particular, we are not aware of any simulations of jet confinement
by an external, supersonic wind.  We naively expect suppression of the instability in cases where the
external wind is mildly relativistic, owing to the small velocity shear.

\section{Summary \& conclusions}\label{summary}

We have constructed a class of semi-analytic models for the collimation and confinement of an accelerating, ultra-relativistic 
outflow by the pressure and inertia of a surrounding medium, under conditions anticipated near the central engines of GRBs.  We 
considered two different scenarios: In the first one the inner flow is assumed to be confined
by the kinetic pressure of a subsonic (or static) corona that extends vertically over several characteristic scales.   
In the second one it is envisaged that the inner outflow collides with a supersonic, baryon rich wind emanating from a disk or a torus surrounding the inner source.  Confinement in this case is accomplished by the ram pressure of the exterior wind.
In general, the collision of the inner jet with the external medium creates a strong oblique shock across which the streamlines of the jet are deflected.  In the case where the inner jet is confined by the ram pressure of a supersonic wind a second shock forms across which the exterior wind gets deflected.  The shocked jet and wind layers are separated by a contact discontinuity surface across which the pressure is continuous.  The gross features described above are observed in numerical simulations (Alloy et al. 2000, 2005), although temporal effects complicate the structure.  The evolution of the shocked jet layer depends, quite generally, on the configuration of the confining medium (the pressure profile in the case of a static corona or the wind geometry in the case of confinement by a supersonic wind), and the  opening angle of the injection cone of the inner outflow.  We identified cases where the inner shock converges towards the axis, and cases where it diverges away from the axis (depending on the opening angle of the inner flow).  In the former case the inner flow quickly becomes fully shocked, whereas in the latter case the structure of the polar outflow consists of a core containing the unshocked ultrarelativistic jet enveloped by the shocked jet layer that expands relativistically, but with a Lorentz factor considerably smaller than that of the unshocked jet. The observational consequences of such a structure have been discussed elsewhere (Eichler and Levinson 2003, 2004; Levinson and Eichler 2004, 2005). 

The structure of the inner jet, when confined by the ram pressure of a supersonic wind, depends on the geometry of the wind and the conditions inside the shocked wind layer.  For wind geometries with radially diverging streamlines (e.g., split monopole) we find that if  
the shocked wind layer can be considered negligibly thin (e.g., due to rapid cooling), then after an initial collimation phase the inner jet  becomes conical with an opening angle that scales roughly as $(L_j/L_w)^{1/3}$ above the geometrical limit, where $L_j$ and $L_w$ are the total power of the unshocked jet and wind, respectively.  A stronger dependence of the opening angle on the power output ratio has been found for exterior wind configurations in which some streamlines are inclined towards the axis, as in the case of a wind from a thin torus invoked in LE00.   In the absence of energy losses from the shocked wind layer the streamlines of the shocked baryon poor flow diverge slower than conical, owing to the larger pressure of the shocked wind.  For the cases examined above we find that the cross sectional radius of the contact discontinuity scales roughly as $r_c\propto z^{2/3}$.  However, at some radius the outer shock and the contact discontinuity become causally disconnected, at which point collimation should be less effective.  The opening angle of the inner jet at the decoupling point is typically smaller than that found in the case of an infinite compression ratio shock.   The formation of a rarefaction wave may act as a booster on the BPJ fluid contained between the contact discontinuity and the rarefaction wave, as demonstrated by Aloy \& Rezzolla 2006, although this effect should be explored further for the situation considered here.  Free neutrons leaking across the interface separating the baryonic wind and the shocked BPJ layer will be picked up by the BPJ, leading to baryon loading of the inner flow and the consequent emission of high-energy neutrinos and $\gamma$-rays, as discussed in Levinson \& Eichler (2003).  However, Kelvin-Helmholtz instabilities at the interface may alter this picture in ways that need to be explored. 

In situations in which the inner jet is confined by the kinetic pressure of a medium having a pressure profile of the form $p\propto z^{-\eta}$, the cross-sectional radius of the shocked baryon poor jet scales as $r_c\propto z^{\eta/4}$.  If the pressure is contributed by a coasting wind, then collimation ($\eta<4$) is expected to occur naturally in the acceleration zone of the inner, baryon poor flow.

It has been argued that internal shocks cannot form in the acceleration zone of a fireball since different fluid shells cannot catch up there.  As demonstrated above, collisions of the accelerating baryon poor outflow with exterior medium can lead to dissipation of a considerable fraction of its bulk energy below the coasting region through formation of strong oblique shocks.
Moreover, reflection of those shocks at the symmetry axis, which may occur under certain conditions, and/or intermittent ejection of the inner outflow and/or the confining wind should lead to large temporal variations of the Lorentz factor of the deflected flow at a given position, and the consequent 
collisions of different blobs of baryon poor fluid that have been deflected by the confining medium (see also Eichler 1994; Aloy et al. 2000).  The internal shocks thereby created would give rise to additional dissipation over a large range of radii.  Those shocks are expected to be radiation mediated.  The radiative efficiency is expected to be very high if the photosphere is located below the coasting region.    
The photon distribution produced inside those shocks is likely to have a nonthermal extension at high energies owing to multiple Compton scattering on the converging flow (a relativistic version of the Blandford Pyne process).  If the shocks are produced at a modest optical depth then the nonthermal photons may escape before being thermalized, giving rise to a nonthermal spectral component of the prompt GRB emission, as proposed earlier (e.g., Eichler 1994; Levinson 2006b, and references therein).  Detailed calculations are required to assess whether this process can indeed explain the observed spectra. 

Finally, we comment on the applicability of this model to other systems. 
Opacity arguments suggest that the rapid variability of the TeV emission observed in the TeV blazars
implies high Doppler factors of the emitting plasma in the $\gamma$-ray emission zone (e.g., Levinson 2006b).
Such high values of the Doppler factor are in clear disagreement with the much lower values inferred from 
radio observations (e.g., Urry \& Padovani 1991; Marscher 1999; Jorstad et al. 2001).
Various explanations, including jet deceleration (Georganopoulos \& Kazanas 2003), a structure 
consisting of interacting spine and sheath (Ghisellini et al. 2005), and opening angle 
effects (Gopal-Krishna et al. 2004) have been proposed in order to resolve this discrepancy. 
Stationary radio features have been observed also in other systems, and on various scales, where relativistic expansion is inferred or expected (e.g., HST1 knot in M87).  We emphasize that, while the Lorentz factors inferred from radio observations reflect the speed of the emission pattern, the rapid variability of the very high energy $\gamma$-ray emission constrains the Lorentz factor of the fluid elements emitting the $\gamma$ rays.  Dissipation behind recollimation shocks can naturally account for large differences between the speeds of the dissipation pattern and the outflowing matter.  In particular, stationary features in a relativistic outflow can be easily produced, as demonstrated by Komissarov \& Falle (1997).  While it is quite likely that the TeV flares in the BL Lac objects are produced by decelerating fronts, recollimation on small scales, particularly in the case of efficient radiative cooling, can alleviate the requirements on the Lorentz factors (Levinson 2007, in preparation).  It has been proposed recently (Stawarz et al. 2007)  
that the HST1 knot in the M87 jet may be associated with a converging shock in a reconfinement nozzle.

\acknowledgements
We thank M. Aloy for a careful reading of the manuscript and for constructive criticism.
This work was supported by an ISF grant for the Israeli Center for High Energy Astrophysics.

\appendix
\section{Derivation of the flow equations}\label{app_A}

We divide the shocked layers into slices of thickness $dz$ along the symmetry axis (see fig 7).  
Equations (\ref{cont-j}) and (\ref{T-j}) are then integrated over a volume of the shocked BPJ enclosed by the inner shock surface, the contact discontinuity surface, and two planes perpendicular to the $z$-axis, located at the points $z$ and $z+dz$.  Using Gauss's theorem we obtain,
\begin{eqnarray}
\left[\int_{r_j}^{r_c}\rho_{js}u^z_{js}2\pi rdr\right]_z^{z+dz} +
\int_{\partial \Omega_j} \rho_{js}u^k_{js}{\hat n}_{jk} dS_j+ \int_{\partial \Omega_c} \rho_{js}u^k_{js}{\hat n}_{ck} dS_c &=& 0\label{app_1},\\
\left[\int_{r_j}^{r_c}T^{\mu z}_{js}2\pi rdr\right]_z^{z+dz} +
\int_{\partial \Omega_j} T^{\mu k}_{js}{\hat n}_{jk} dS_j+ \int_{\partial \Omega_c} T^{\mu k}_{js}{\hat n}_{ck} dS_c &=& 0\label{app_2},
\end{eqnarray}
where ${\hat n}_{j}$ and ${\hat n}_{c}$ are the normals to the inner shock and contact discontinuity surfaces, respectively,
${\partial \Omega_j}$ and ${\partial \Omega_c}$ denote integration over a portion of the shock and contact discontinuity surfaces enclosed between $z$ and $z+dz$, respectively, and $dS_j$ and  $dS_c$ are the corresponding surface area elements, as indicated in fig 7.  The latter are given explicitly by
\begin{equation}
dS_{j(c)} = 2\pi\frac{r_{j(c)}d z}{\cos\alpha_{j(c)}}, \label{app_dS}
\end{equation}
where the angles $\alpha_j$ and $\alpha_c$ are related to the cross-sectional radii $r_j(z)$ and $r_c(z)$ through: $\tan\alpha_{j(c)}=dr_{j(c)}/dz$. 
Recalling that the BPJ streamlines are tangent to the contact discontinuity surface we have $u^k_{js}{\hat n}_{ck}=T^{0k}_{js}{\hat n}_{ck}=0$ and $T^{zk}_{js}{\hat n}_{ck}=p_{js}\sin\alpha_c$.
Now, the energy flux $T_{js}^{0z}=w_{js}\gamma_{js}u^z_{js}$ involves the poloidal velocity $u_{js}$ that appears in the Lorentz factor  $\gamma_{js}=(u_{js}^2+1)^{1/2}$, which on any given streamline of the shocked BPJ fluid is given 
in terms of the angle between the direction of the streamline and the axis, $\psi(r,z)$, as:
$u^z_{js}=u_{js}\cos\psi$.
Since $\psi\le\alpha_c$ we have  $2(1-u^{z}_{js}/u_{js})<\alpha_c^2<(dr_c/dz)^2$ at small angles. To order $(dr_c/dz)^2$ we can therefore 
replace $u^z_j$ by $u_j$ thereby eliminating one variable.
Under the assumption of our 1D model, that the flow quantities depend only on the coordinate $z$, the above results yield,
\begin{eqnarray}
\frac{d}{dz}\left[\rho_{js}u_{js}(r_c^2-r_j^2)\right]+
\left[2\rho_{js}u^k_{js}{\hat n}_{jk}\frac{r_j}{\cos\alpha_j}\right]_{r=r_j(z)}&=& 0\label{app_3},\\
\frac{d}{dz}\left[T^{\mu z}_{js}(r_c^2-r_j^2)\right]+
\left[2T^{\mu k}_{js}{\hat n}_{jk}\frac{r_j}{\cos\alpha_j}\right]_{r=r_j(z)}+p_{js}\sin\alpha_c g^{\mu z}
&=& 0\label{app_4}.
\end{eqnarray}
By employing the jump conditions at the shock front and the contact discontinuity surface (eqs. [\ref{jump-js1}] and [\ref{jump-js2}]), 
and using eqs.  (\ref{def-delta1}) and (\ref{def-delta2}) we finally arrive at eqs (\ref{mass_eq_j}), (\ref{energy_eq_j}) and (\ref{momZ_eq_j}).
Likewise, eqs. (\ref{cont-w}) and (\ref{T-w}) can be integrated over the volume of the 
shocked wind enclosed between $z$ and $z+dz$.  Using the same procedure as above we obtain eqs. (\ref{mass_eq_w}), (\ref{energy_eq_w}) and (\ref{momZ_eq_w})

Two additional relations can be obtained by projecting the jump conditions (\ref{jump-js1}) and (\ref{jump-ws1}) on the directions normal to the inner and outer shock surfaces, respectively, using eqs. (\ref{T_M}) and (\ref{p-flux-jn}): 
\begin{eqnarray}
p_{js}+w_{js}(u_{js}^k{\hat n}_{jk})^2&=& p_{j}+w_{j}u_{j}^2\sin^2\delta_{j},\label{app_5}\\
p_{ws}+w_{ws}(u_{ws}^k{\hat n}_{wk})^2&=& p_{w}+w_{w}u_{w}^2\sin^2\delta_{w}\label{app_6}.
\end{eqnarray}
The terms associated with the ram pressure of the downstream fluid (the second terms on the left-hand side of eqs. [\ref{app_5}] and [\ref{app_6}]) depend on the direction of deflected streamlines just behind the shock surfaces, which cannot be computed under the assumptions of our model.  However, these terms are typically smaller than the post shock kinetic pressure.  For a planar shock $w_{s}(u_{s}^k{\hat n}_{k})^2/p_{s}=1/6$ for a non-relativistic, radiation dominated shock, and $w_{s}(u_{s}^k{\hat n}_{k})^2/p_{s}=1/2$ in the extreme relativistic limit.  For an oblique shock this ratio is even smaller.  Thus, those terms can be neglected, whereby eqs. (\ref{momN3_eq_j}) and (\ref{momN3t_eq_w}) are obtained.

\clearpage
\begin{figure}[h]
\centering
\includegraphics[width=11.0cm]{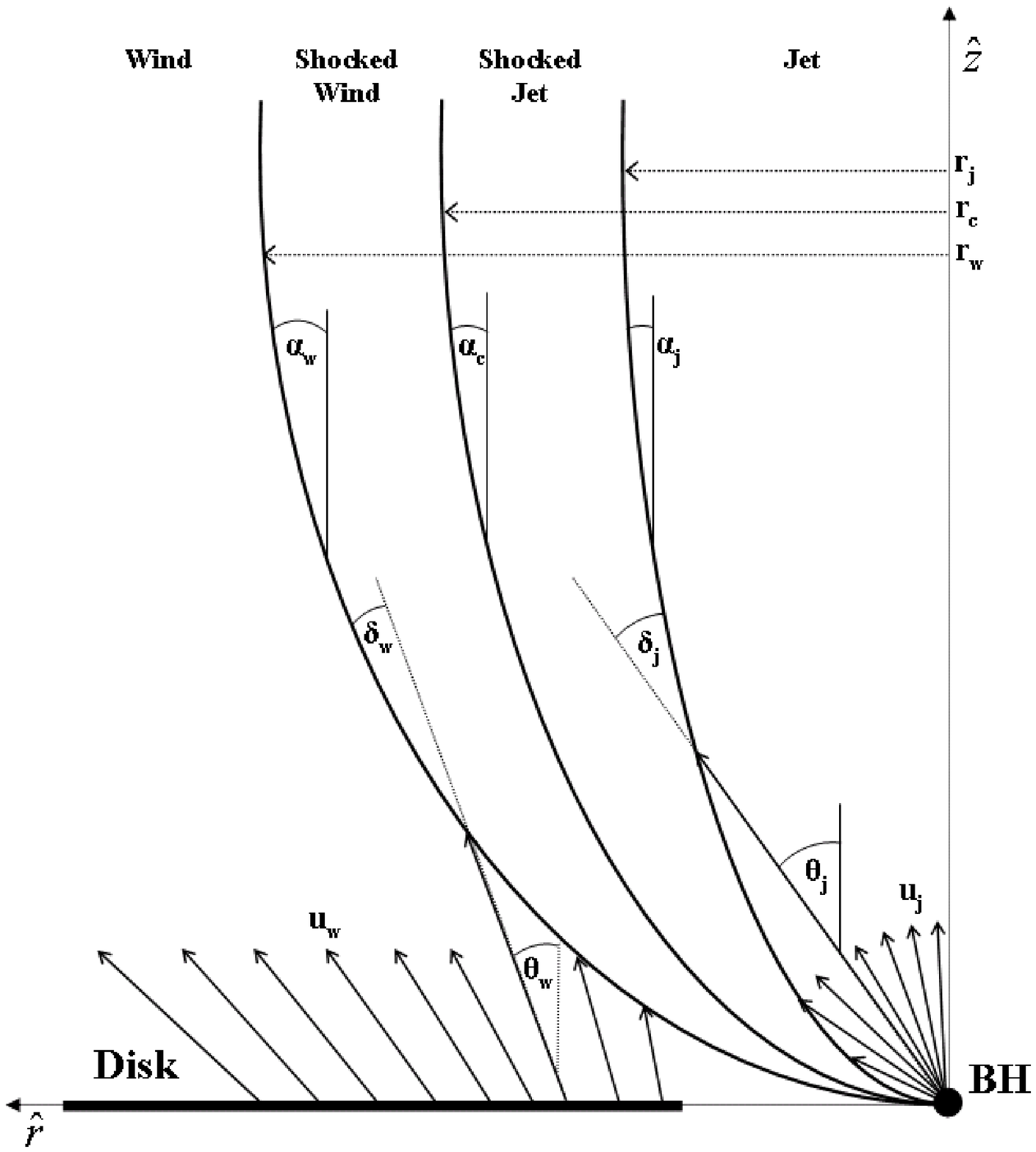}
\label{fig.jet}
\caption{\small{A sketch of the colliding outflows model.  The
unshocked BPJ, shocked BPJ, unshocked wind, and shocked wind zones are designated as $j$, $js$,
$w$, and $ws$, respectively. }}
\end{figure} 
%

\begin{figure}[h]
\centering
\includegraphics[width=11.0cm]{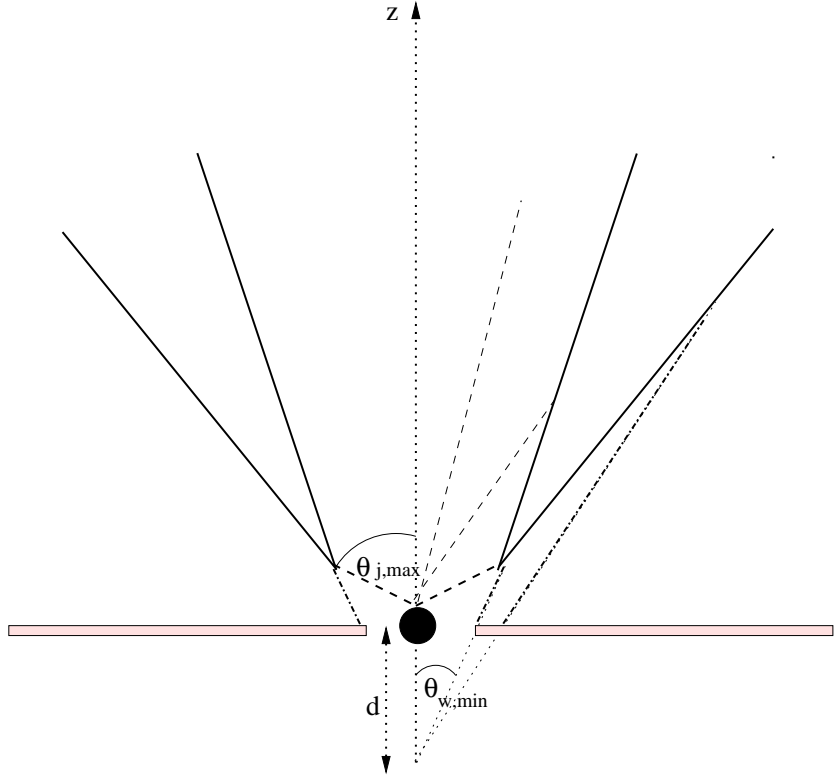}
\label{fig-split}
\caption{\small{A sketch of the split monopole wind geometry.  The streamlines of the unshocked wind are represented by the 
dot-dashed lines. The minimum angle of the wind $\theta_{w,min}$ is also indicated.  
The dashed and solid lines mark the streamlines of the unshocked BPJ and the inner and outer shock surfaces, respectively.
The BPJ is assumed to be injected into a cone of opening angle $\theta_{j,max}$, as indicated.}}
\end{figure} 

\begin{figure}[h]
\centering
\includegraphics[width=12cm]{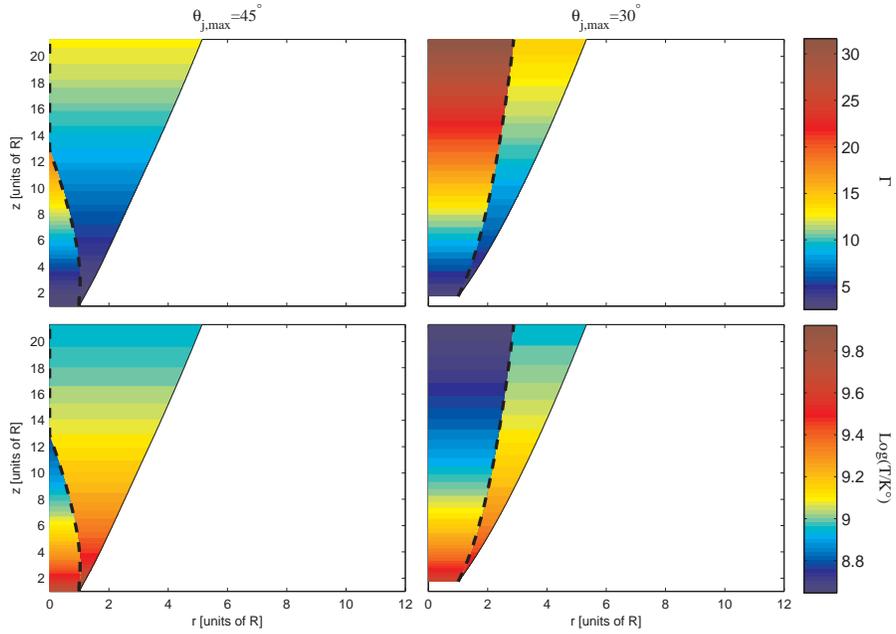}\label{press_v}
\label{fig:press_v}
\caption{\small{Lorentz factor (top) and logarithm of the temperature (bottom) of the shocked and unshocked BPJ in the case of confinement by a static corona with a pressure profile $p_w\propto z^{-3}$, for two different angles of the BPJ injection cone, as indicated. The dashed line marks the inner shock surface.}}
\end{figure}

\begin{figure}[h]
\centering
\includegraphics[width=12cm]{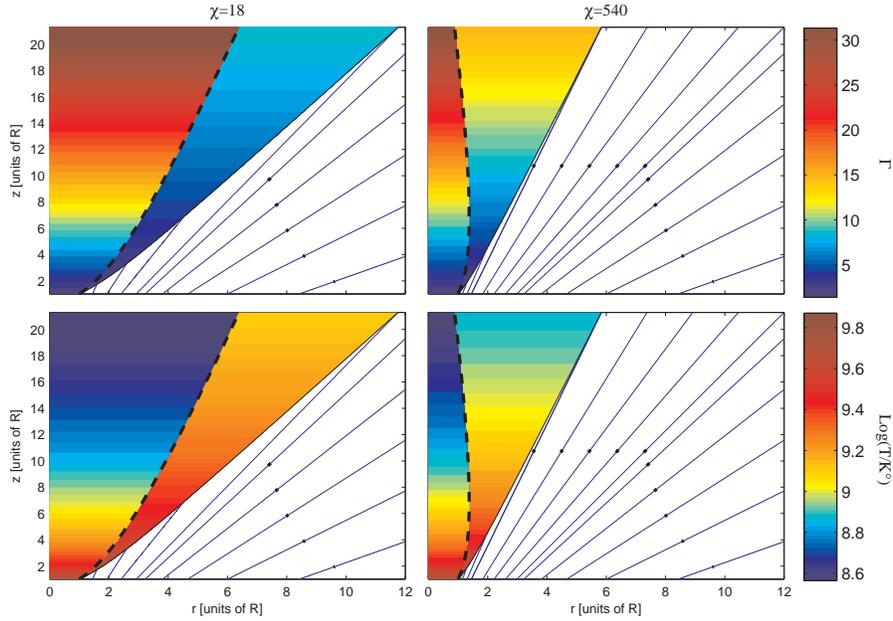}    
\label{fig:ram_point_v}
\caption{\small{Lorentz factor (top) and logarithm of the temperature (bottom) of the shocked and unshocked BPJ in the case of confinement by a split monopole wind with $\theta_{w,min}=8.5^{\circ}$, for two values of the luminosity ratio $\chi$, as indicated.  In both cases shown  $L_j/4\pi c R^2=2\times 10^{26}$ erg cm$^{-3}$, where $R=d\tan\theta_{w,min}$ is the radius at which the innermost streamline of the wind intersects the equatorial plane.
The blue solid lines represent the streamlines of the unshocked wind.  The shocked wind layer is assumed to be negligibly thin in this calculation (that is, the outer shock front coincides with the contact discontinuity surface). The opening angle of the BPJ injection cone in both cases shown is $\theta_{j,max}=45^\circ$.  The opening angle of the shocked BPJ (specifically, the contact discontinuity surface) approaches $27^\circ$ and $13^\circ$ at $z/R\sim$ a few for $\chi=18$ and $\chi=540$, respectively.}}
\end{figure} 

\begin{figure}[h]
\centering
\includegraphics[width=12.0cm]{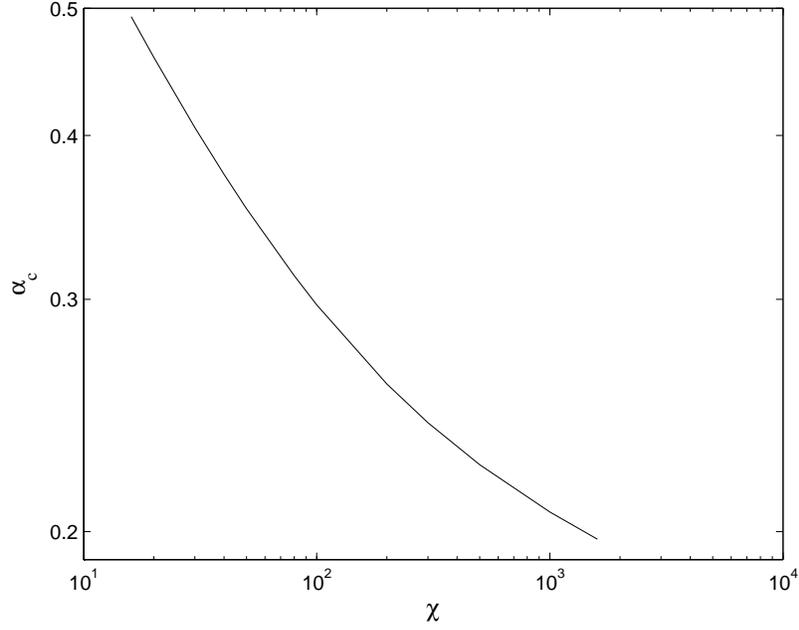}
\label{alph-chi}
\caption{\small{Dependence of $\alpha_{c0}$ on $\chi$, computed for a split monopole wind with $\theta_{w,min}=8.5^\circ$.}}
\end{figure}

\begin{figure}[h]
\centering
\includegraphics[width=12cm]{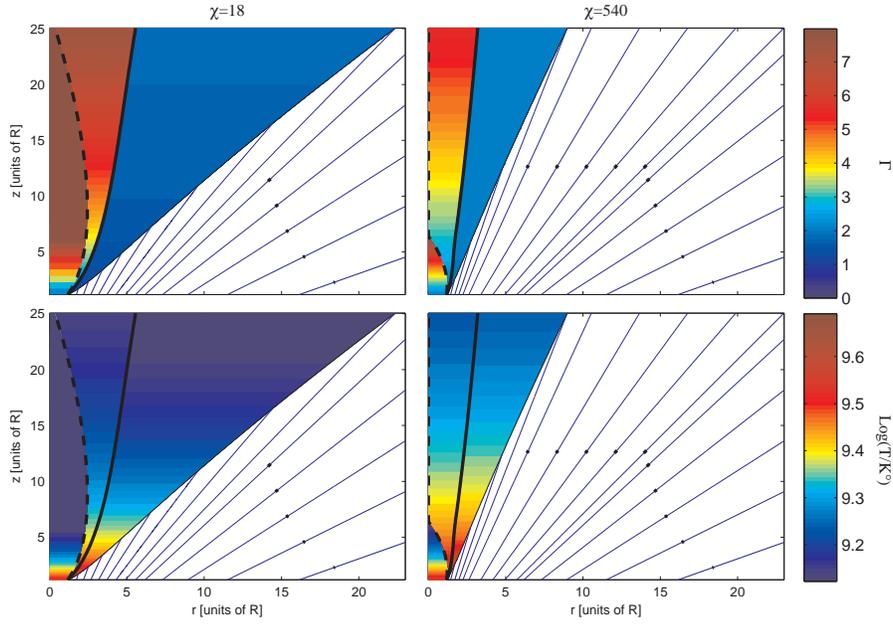}
\label{fig:ram_point_t}
\caption{\small Same as fig 4, but with no energy losses from the shocked wind layer.  The parameters of the split monopole wind and the angle $\theta_{j,max}$ are the same as in fig 4. The inner shock is marked by the dashed line and the contact discontinuity surface by the thick solid line.  The colored region to the right of the thick solid line corresponds to the shocked wind layer. }
\end{figure}

\begin{figure}[h]
\centering
\includegraphics[width=14cm]{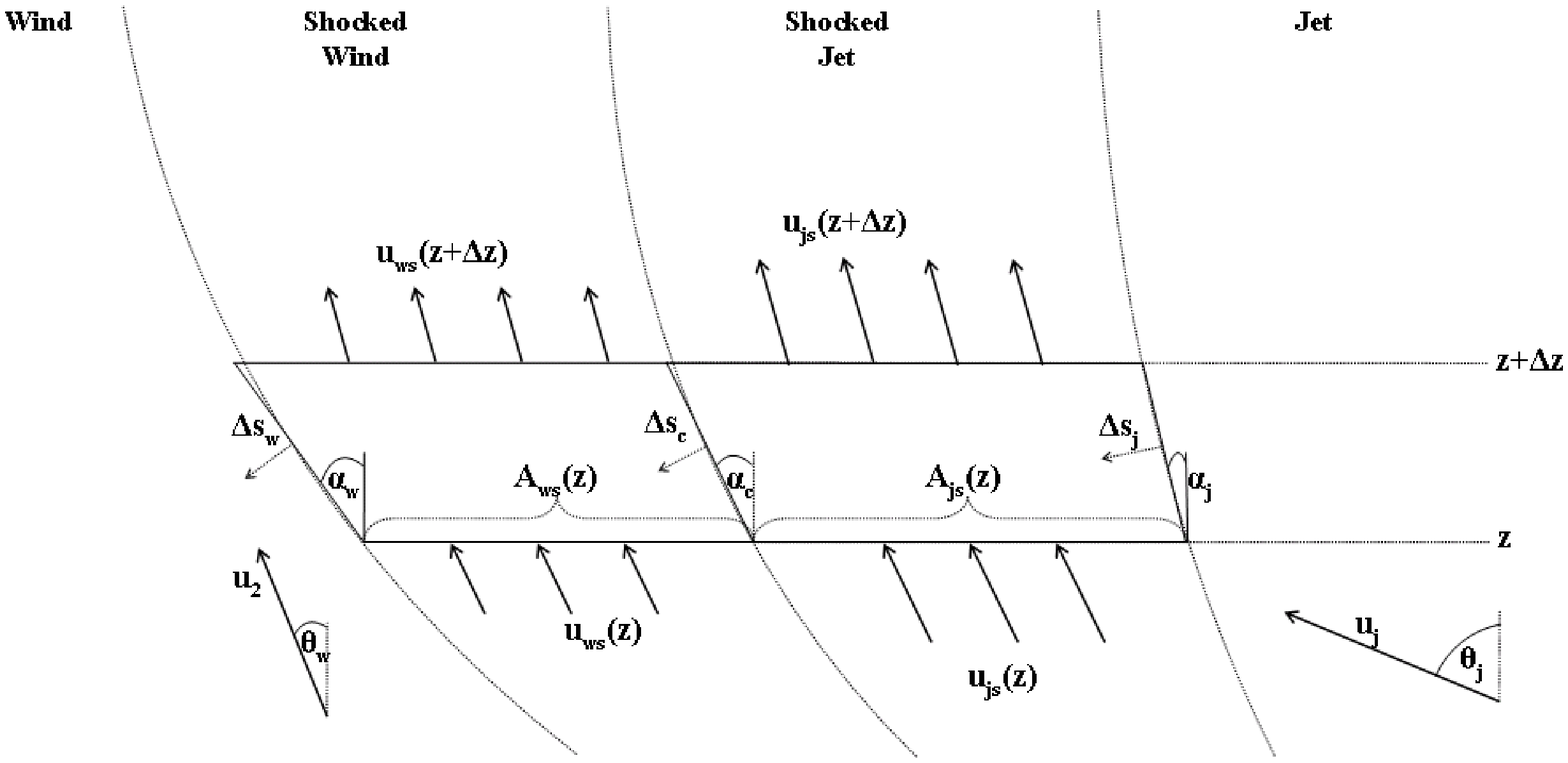}
\caption{\small{Schematic diagram showing the slice of shocked BPJ and wind layers, and the corresponding contours of integration.
The various quantities used in the derivation of the model equations are indicated}}
\end{figure} 
%


\begin{thebibliography}{9}
\bibitem[]{2000ApJ..L119} Aloy, M.A., Muller, E., Ibanez, J.M., Marti, J.M. \& MacFadyen, A. 2000, ApJ, 531, L119
\bibitem[]{Aloy05} Aloy, M.A., Janka, H-T. \& Muller, E. 2005, A\&A, 436, 273
\bibitem[]{Aloy06} Aloy, M.A. \& Rezzolla, 2006, ApJ, 640, L115
\bibitem[]{Beegel1994} Begelman, M. \& Li, Z-Y. 1994, ApJ, 426, 269
\bibitem[]{Beegel1998} Begelman, M. 1998, ApJ, 493, 291
\bibitem[]{2006MNRAS..375} Beskin, V.S. \& Nokhrina, E.E. 2006, MNRAS, 367, 375
\bibitem[]{2005Sci..1833} Burrows, D.N. et al. 2005, Science, 309, 1833
\bibitem[]{1982ApJ..571} Eichler, D. 1982, ApJ, 263, 571
\bibitem[]{1993ApJ..419} Eichler, D. 1993, ApJ, 419, 111
\bibitem[]{1994ApJ..419} Eichler, D. 1994, ApJS, 90, 877
\bibitem[]{2003ApJ..L147} Eichler, D. \& Levinson, A. 2003, ApJ, 596, L147
\bibitem[]{2004ApJ..L13} Eichler, D. \& Levinson, A. 2004, ApJ, 614, L13
\bibitem[]{2005MNRAS..L42} Fan, Y.Z. \& Wei, D.M. 2005, MNRAS, 364, L42
\bibitem[]{590} Georganopoulos M. and Kazanas, D. 2003, ApJ, 594, L27
\bibitem[]{591} Ghisellini, G. Tavecchio, F. and Chiaberg, M. 2005, A\&A, 432, 401
\bibitem{Gopal-Krishna04} Gopal-Krishna, S. Dhurde, S. \& Wiita. P.~J. 2004, New Astron.,615, L81
\bibitem[]{2006MNRAS..1946} Granot, J., Konigle, A. \& Piran, T. MNRAS, 370, 1946
\bibitem[]{2003ApJ..583} Hardee, P.~E. \& Hughes, P.~A. 2003, ApJ, 583, 116  
\bibitem[]{2006ApJ..103} Hawley, J. F. \& Krolik, J. 2006, ApJ, 641, 103
\bibitem{Jorstad01} Jorstad, S., Marscher, A.P., Mattox, J.R., Werhle, A.E., Bloom, S.D.,
and Yurchenko A.V. 2001, ApJ, 134, 181
\bibitem[]{2007RMAA..1_01} Krolik, J. 2007, RevMexAA, 27, 1
\bibitem[]{2007RMAA..1_02} Komissarov, S.~S. \& Falle, S.~A. 1997, MNRAS, 288, 833
\bibitem[]{1993ApJ..418} Levinson, A., \& Eichler, D. 1993, ApJ, 418, 386
\bibitem[]{2000PRL..236} Levinson, A., \& Eichler, D. 2000, PRL, 85, 236
\bibitem[]{2004ApJ..L19} Levinson, A., \& Eichler, D. 2003, ApJ, 594, L19
\bibitem[]{2004ApJ..1079} Levinson, A., \& Eichler, D. 2004, ApJ, 613, 1079
\bibitem[]{2005ApJ..L13} Levinson, A., \& Eichler, D. 2005, ApJ, 629, L13
\bibitem[]{2006IJMPA..510} Levinson, A. 2006a, ApJ, 648, 510
\bibitem[]{2006IJMPA..6015} Levinson, A. 2006b, IJMPA, 21, 6015
\bibitem[]{1992SAL..878} Lyubarsky, Y. 1992, Sov. Astron. Lett., 18, 878
\bibitem[]{2006MNRAS..1594} Lyutikov, M. 2006, MNRAS, 367, 1594
\bibitem[]{1999ApJ..262} MacFadyen, A. \& Woosley, S.E. 1999, ApJ, 524, 262
\bibitem[]{610} Marscher, A.~P. 1999, Astropart. Phys., 11, 19 
\bibitem[]{2004ApJ..977} McKinney, J. C. \& Gammie, C.F. 2004, ApJ, 611, 977
\bibitem[]{2005ApJ..L5} McKinney, J. C. 2005, ApJ, 630, L5 
\bibitem[]{2006MNRAS..1561} McKinney, J. C. 2006, MNRAS, 368, 1561 
\bibitem[]{1997ApJ..505} Meszaros, P. \& Rees, M.J. 1997, ApJ, 482, 29
\bibitem[]{1986ApJ..L43} Paczynski, B. 1986, ApJ, 308, L43  
\bibitem[]{2005RMP..1143} Piran T. 2005, Rev. Mod. Phys., 76, 1143
\bibitem[]{1999ApJ..518} Popham, R. Woosley, S.E., \& Fryer, F., 1999, ApJ, 518, 356
\bibitem[]{2003ApJ..L5} Proga, D. MacFadyen, A.I., Armitage, P.J. \& Begelman, M.C. 2003, ApJ, 599, L5
\bibitem[]{Pruet03} Pruet, J. Woosley, S.E., \& Hoffman, R.D. 2003, ApJ, 586, 1254
\bibitem[]{2003FM..199} Rezzolla, L., Zanotti, O. \& Pons, J.A. 2003, J. Fluid Mech., 479, 199
\bibitem[]{2003MN..L36} Rosswog, S. \& Ramirez-Ruiz, E. 2003, MNRAS,343, L36
\bibitem[]{622} Stawarz, L., et al. 2006, MNRAS, 370, 981
\bibitem{Urry91} Urry, C.M., and Padovani, P. 1991, ApJ, 371, 60
\bibitem[]{2001PRt..1} Van Putten, M.V.P. 2001, Phys. Rep. 345, 1 
\bibitem[]{2003ApJ..1080} Vlahakis, N. \& Konigl, A. 2003, ApJ, 596, 1080
\end{thebibliography}
\end{document}